\documentclass[10pt,letter,twocolumn,twoside,openleft]{computer}

\usepackage[utf8]{inputenc}
\usepackage[english]{babel}
\usepackage{subfig}
\usepackage{comment}             

\title{Improving Hyperconnected Logistics with Blockchains and Smart Contracts}
\date{}

\usepackage{tikz}
\usepackage{cite}
\usepackage{verbatim}
\usepackage[export]{adjustbox}
\usepackage{listings}
\usepackage{xcolor}
\PassOptionsToPackage{hyphens}{url}\usepackage{hyperref}
\usepackage{paralist}

\usepackage{todonotes}

\definecolor{dkgreen}{rgb}{0,0.6,0}
\definecolor{gray}{rgb}{0.5,0.5,0.5}
\definecolor{mauve}{rgb}{0.58,0,0.82}
\definecolor{cmdline}{HTML}{e6e6e6}

\lstdefinestyle{javacode}{
    frame=tb,
    language=Java,
    aboveskip=3mm,
    belowskip=3mm,
    showstringspaces=false,
    columns=flexible,
    basicstyle={\small\ttfamily},
    numbers=left,
    numberstyle=\small\color{gray},
    keywordstyle=\color{blue},
    commentstyle=\color{dkgreen},
    stringstyle=\color{mauve},
    breaklines=true,
    breakatwhitespace=true,
    tabsize=3
}

\lstdefinestyle{console}{
    basicstyle=\linespread{1.1}\ttfamily\footnotesize,
    backgroundcolor=\color{cmdline}
}

\usepackage[htt]{hyphenat}

\usepackage{paralist}
\usepackage{csquotes}
\usepackage{subfig}

\usetikzlibrary{arrows,shapes}

\newcommand{\etal}{\textit{et al}.~}
\newcommand{\ie}{\textit{i}.\textit{e}.}
\newcommand{\eg}{\textit{e}.\textit{g}.}

\newcommand{\SCShipment}{\emph{Shipment}}
\newcommand{\SCShipmentManager}{\emph{ShipmentManager}}

\graphicspath{{fig/}}

\begin{document}

  \twocolumn[
  \begin{@twocolumnfalse}
  \maketitle

  {\large\it Quentin Betti, Raphaël Khoury, Sylvain Hallé}\\
  {\small Université du Québec à Chicoutimi, Canada}
  
  \vskip 0.2cm
  
  {\large\it Benoit Montreuil}\\
  {\small Georgia Institute of Technology, USA}
  
  \vskip 0.8cm
  \colorbox{MyGray}{\footnotesize\rule{2in}{0in}}
  
  \noindent \begin{flushleft}{\fontfamily{phv}\fontseries{b}\selectfont\colorscheme
The Physical Internet and hyperconnected logistics concepts promise an open, more efficient and environmentally friendly supply chain for goods. Blockchain and Internet of Things technologies are increasingly regarded as main enablers of improvements in this domain. We describe how blockchain and smart contracts present the potential of being applied to hyperconnected logistics by showing a concrete example of its implementation. 

  }\end{flushleft}
  \vskip 16pt
  \end{@twocolumnfalse}
  ]


\lettrine{S}upply chain logistics is a highly competitive and legally sensitive industry, where several protagonists with diverse capabilities and goals have to cooperate to fulfill each customer demand. The different actors involved do not necessarily trust each other, and therefore rarely share their goods handling or IT infrastructure; this has been recognized as a major cause of inefficiency \cite{PI}. The concept of \emph{Physical Internet} \cite{PI} is an emerging paradigm that aims to increase supply chain efficiency in terms of goods handling, routing and storage, as well as to improve its economical, environmental and social sustainability.

In order to fulfill this promise, the Physical Internet takes advantage of disruptive new technologies. For example, much prior research points out growing interest in the Internet of Things (IoT) as a technology that can improve the efficiency of the supply chain \cite{Sallez2016,Li2011,Kelepouris2007,Atzori2010, morley2016}. In addition, blockchain technologies and their implementations \cite{Halle2018, Tian2017,Toyoda2017,Korpela,Madhwal2017,Lu2017} make possible for all intervening parties to share an open and trustworthy information system, 
thus reducing the risk of failures or fraud. 

The integration of these various technologies into a cohesive system, however, still remains an open challenge. In the following, we present recent progress made on the application of blockchain and smart contracts to the Physical Internet, and hyperconnected logistics in particular, via a concrete implementation to an existing simulation designed with AnyLogic, a widely used simulation platform.

\section*{A Simple Hyperconnected Logistics Scenario}\label{sec:initial_simulation}

%

\begin{figure*}
    \vspace{0.25cm}
    \centering
    \includegraphics[width=0.75\paperwidth,center]{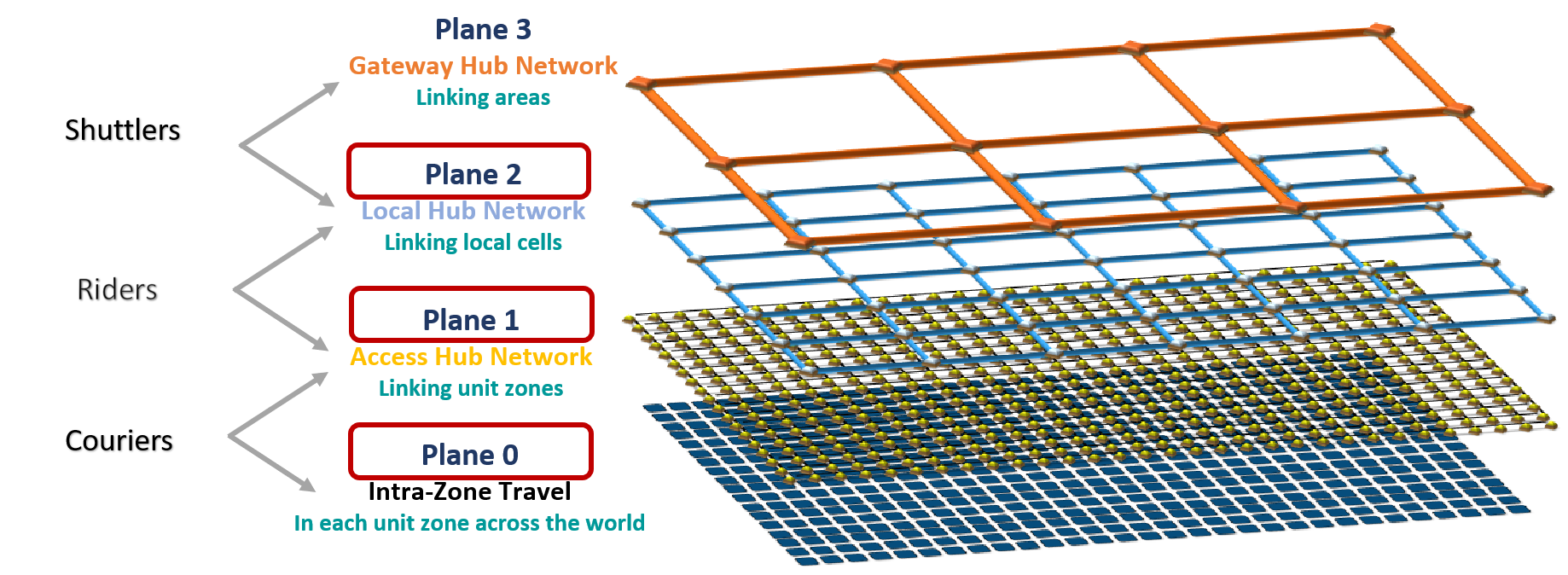}
    \caption{Urban parcel logistic web from Montreuil \cite{Montreuil2018}.}
    \label{fig:hl_web}
    \vspace{0.25cm}
\end{figure*}

The hyperconnected logistics paradigm \cite{Montreuil2018, crainic2016} aims, among other things, at improving the efficiency of goods delivery in terms of package routing, delivery speed and inventory management. This is achieved by evolving away from a hub-and-spoke architecture towards so-called \emph{hyperconnected} networks \cite{Montreuil2018}. In this model, shown in Figure~\ref{fig:hl_web}, the entire world can be split at the smallest scale into \emph{unit zones}, whose size depends on expected demand density. Adjacent unit zones are grouped into local \emph{cells}, which in turn are gathered into \emph{areas}, which form \emph{regions}. Simultaneously, several hub networks are defined to link these different layers: \emph{access hubs} link unit zones together; \emph{local hubs} link local cells, and \emph{gateway hubs} link areas. Different hub levels may exist inside the same physical entity (\eg, a local hub might also be an access hub), thus allowing interactions between the different layers.

\begin{figure*}
    \vspace{0.25cm}
    \centering
    \includegraphics[width=0.75\paperwidth,center]{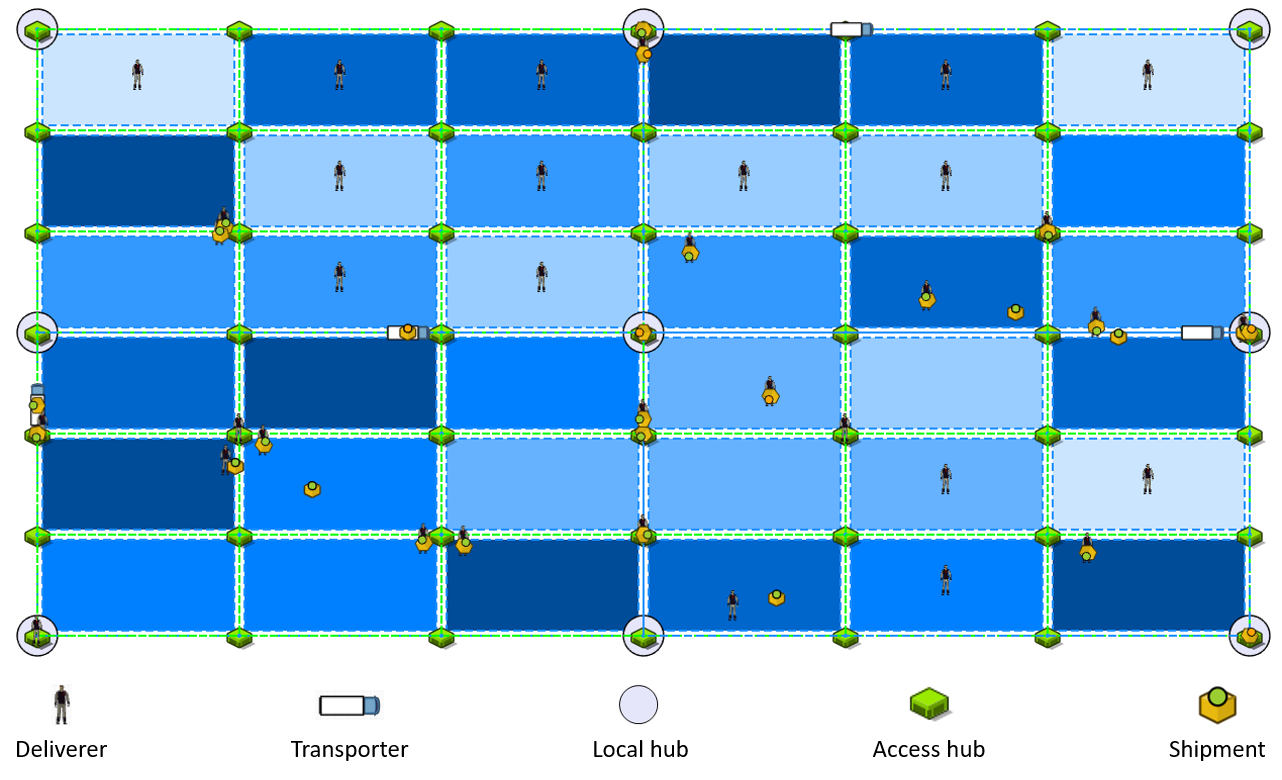}
    \caption{Representation of the megacity and its components \cite{Kaboudvand2018}.}
    \label{fig:sim_map}
    \vspace{0.25cm}
\end{figure*}

To illustrate the potential of the architecture on supply chain efficiency, consider the following simple model of a hyperconnected ``megacity''. The city itself is  modeled as a rectangular map and actually limited to one area (see Figure \ref{fig:sim_map}); this area is itself composed of four local cells. Shipments to be delivered pop up across the city, and both transporters and deliverers participate in shipping them to their final destination inside the city. Each local cell is divided in nine unit zones, whose brightness reflects their demand density: the darker a unit zone's color is, the higher the probability that the zone will be the origin or the destination of a shipment request. Access hubs are located at each vertex of a unit zone and are shared with their adjacent unit zones. Local hubs are located at each vertex of a local cell and are shared by their adjacent local cells. In this scenario, local hubs are also access hubs.

Three kinds of carriers are defined in this simulation: \emph{couriers}, \emph{riders} and \emph{shuttlers}. Couriers take care of last-mile deliveries within their assigned unit zone, doing so walking, biking, or yet riding a scooter or light electrical vehicle. Riders and shuttlers use faster vehicles (usually electrical or natural gas) as they cover longer distances. Riders are assigned to a local cell, moving goods between unit zones and/or local hubs within the cell. Shuttlers are assigned to an area, moving goods between local hubs and/or gateway hubs within the area. Shuttlers preferably use faster roadways than riders, similarly as riders preferably use faster roadways than couriers.


A simulation model of this hyperconnected urban logistic was developed by Kaboudvand \etal with the AnyLogic software and presented at the last IISE Annual Conferenc e\cite{Kaboudvand2018}. In this simulation, riders and shuttlers are gathered in a group of carriers called \emph{tranporters}.

\section*{Blockchain-enabled tracking system in simulation}\label{sec:simulation}


Although the initial simulation \cite{Kaboudvand2018} succeeded in showing that hyperconnected networks are way more efficient than traditional hub-and-spoke networks from an optimization perspective, it does not address the topic of shipments' tracking data management and storage.
Recent works tackled this issue by pointing out the applicability of blockchain networks to hyperconnected environments \cite{hofman2017, hofman2017prez, hofman2019}.  
This section, however, presents a concrete real-time tracking system of shipments, applied to the initial simulation by using an Ethereum blockchain backend. Ethereum (ETH) was the first platform to enable the deployment of smart contracts on the blockchain, is open-source and benefits from a huge community, making it easy to find documentation on how to deal with smart contracts.

\subsection*{Purpose}\label{subsec:simulation_purpose}

The aim of this work is to provide tracking information about actions performed by agents on shipments in the simulation, and these pieces of information must be stored using a blockchain system, instead of regular databases or log files. 

In this simulation, we do not take into account the processing of shipments inside a hub, meaning that the only actions being tracked are shipment deliveries and pick-ups by couriers and transporters. These entities are the \emph{agents} of our simulation. Therefore, in a blockchain perspective, each agent has an account and is capable of making transactions.

The pieces of information we store in blockchain transactions are \emph{actions}, and they are composed as follows:
\begin{inparaenum}
    \item the identifier of the shipment it is performed on;
    \item the date and time it has occurred;
    \item its type (in this scenario, the only types are pick up and delivery);
    \item the name and the account address of the agent performing the action;
    \item its location (\ie, X-Y coordinates in the simulation zone).
\end{inparaenum}

\subsection*{Private blockchain network}

In this model, we simulate a private Ethereum blockchain network which uses a Proof-of-Authority consensus (PoA). This means that we run several Ethereum nodes communicating with each other, and that the blockchain we deal with is completely independent of other networks.

In order to simulate a full-scale network, we chose to assign a full Ethereum client to each agent of the AnyLogic simulation. This means that each courier and transporter receives all the transactions of the network, validates them upon reception, stores them, and thus keeps the whole blockchain locally.
Additionally, we also run \emph{sealers}, which are network nodes responsible for transaction verification and block generation when using a PoA consensus. 

\subsection*{Smart contracts}

As mentioned before, every action is to be stored in the blockchain. To do so, we use two kinds of smart contracts: \SCShipmentManager{} and \SCShipment{} contracts.

\paragraph{Shipment contract}
A \SCShipment{} contract represents a single shipment in the simulation and stores every action performed on it. Thus, it contains the following elements:

\begin{itemize}
    \item \emph{id}: the shipment's identifier (\ie, the name of the shipment in the Anylogic simulation).
    \item \emph{actions}: the list of actions performed on the shipment.
    \item \emph{shipmentManager}: a reference to its \SCShipmentManager{}, whose purpose will be explained later.
\end{itemize}

A \SCShipment{} also provides three callable methods:

\begin{itemize}
    \item \emph{addAction(action)} adds the specified action to the action list. It is only callable by the \SCShipmentManager{}.
    
    \item \emph{getActionCount()} returns the current length of the action list.
    
    \item \emph{getAction(index)} returns the action using its index in \verb+actions+.
\end{itemize}

\paragraph{ShipmentManager contract} 
A \SCShipmentManager{} provides an interface responsible for managing \SCShipment{} contracts that characterize shipments of a same application or context. It is composed of a single attribute and two methods:

\begin{itemize}
    \item \emph{shipmentsById}: a structure mapping shipment ids to their corresponding \SCShipment{} contracts. Therefore, it keeps a reference to all the  \SCShipment{} contracts of a specific set.
    
    \item \emph{addAction(shipmentId, action)} calls the \emph{addAction} method of the \SCShipment{} contract corresponding to the provided \emph{shipmentId} inside \emph{shipmentsById}. If the \SCShipment{} contract does not yet exist, it is created and then the action is added. Moreover, a log event describing the new action is emitted. 
    
    \item \emph{getShipment(shipmentId)} returns the address of the \SCShipment{} contract corresponding to the provided \emph{shipmentId} inside \emph{shimentsById}, or the null address (\ie, \emph{0x0}) if it does not exist.
\end{itemize}

One could argue that there is no need for \SCShipmentManager{} contracts. Indeed, actions could be directly sent to corresponding \SCShipment{} contracts and the result would be the same. However, there are several justifications for this choice. 

First, using only \SCShipment{} contracts implies that the agents must know all of their addresses, 
meaning that some kind of database or file containing these addresses must be maintained and made available to all the agents, who must in turn be able to update it each time they deploy a new contract. A \SCShipmentManager{} contract solves this issue by keeping a reference to every contract and updating this list every time an action is added: there is no need for an external system and the \SCShipmentManager{}'s address is all the agents need to know.

Second, having such an interface allows us to emit events on newly added actions for all shipments on a single contract. Otherwise, each \SCShipment{} contract would be responsible for emitting their own events. If one wanted to monitor the actions realized on specific shipments, he would have to do it on a per-shipment basis. In our solution, he only needs to monitor the events emitted by the \SCShipmentManager{} and filter those deemed interesting.

\subsection*{Global solution}

So far, we have described the blockchain network we use and how actions are stored in the blockchain using smart contracts. This section shows how an agent inside the AnyLogic simulation is able to interact with them.




\paragraph*{Action Managers and architecture}\label{subsubsec:am-arch}

Since AnyLogic simulations are developed in Java and does not provide a framework to handle blockchain technologies, the first idea was to import web3j --- an open-source Java library that allows communications with ETH clients (\url{https://web3j.io/}) --- in AnyLogic and deal with smart contracts directly inside the simulation.
However, there is a conflict between the cryptographic library already present in AnyLogic and the one used by web3j to sign transactions, making this option impossible. Thus, the alternative was to develop a program outside AnyLogic, which would be able to receive actions sent by the AnyLogic agents, then forward them as transactions to the ETH clients. 

This intermediary program is called an \emph{Action Manager}: it is actually an HTTP server aware of the agent's public/private keys, which is listening for POST requests containing action data from the simulation agent, then processes them into actual ETH transactions, and then send them to the ETH client via RPC. This means that there are three different entities at the agent level: \begin{inparaenum}[1)]
\item the agent inside the AnyLogic simulation,
\item its Action Manager responsible for transaction forwarding, and
\item the ETH node which receives transaction from the Action Manager and communicates with the other ETH nodes of the blockchain network.
\end{inparaenum}

\paragraph*{Switching to go-ethereum}

In order to reduce Action Managers' memory consumption (each one of them took up to 200 MB of memory and 40 must be run simultaneously --- for 36 couriers and 4 transporters), it was decided to switch from Java to Go programs. Indeed, the Go language offers very light HTTP requests handling, is the native development language of the Ethereum clients and thus benefits from the same features as web3j. Therefore, Action Managers were re-coded in Go and used the go-ethereum library (\url{https://github.com/ethereum/go-ethereum}), the native official library for Ethereum clients. This change provided highly positive results, since running all the 40 Action Managers now only requires about 20 MB of RAM, against approximately 8GB with Java and web3j.  

\section*{Running the simulation}\label{sec:running}



We developed a Go program using the go-ethereum library to automatically set up and start our network and create the different accounts for our sealers/agents. A \emph{bootnode} \cite{bootnodes} is also started to enable the communication between nodes.

Once the network is up and running, AnyLogic agents need to send their shipment actions to the blockchain. As mentioned in the previous section, Action Managers are required for sending actions from AnyLogic agents to the Ethereum clients. Thus, they must first be started, using our custom Go program, which creates and runs an Action Manager for each of the AnyLogic agents.

At this point, the AnyLogic simulation can be launched. As a result, the Go program displays the output of the first sealer in our simulation, \emph{sealer0}. Obviously, it takes some time for the first transaction to occur if the AnyLogic simulation runs at initial speed. 
This is because of a delay between the beginning of the simulation and the actual first pick up action from one of the deliverer. 
After a few seconds though, nearly every block import should contain at least one transaction (and probably many more) due to the high density of actions in the simulation.


Even though watching new block imports in the program's output is a good way to know if our system is working properly or not at all, listing all added actions in real-time is even better. To this end, an action monitor was also implemented. 
It connects to the \emph{sealer0}'s \SCShipmentManager{} contract, prints \emph{old} actions (\ie, actions that were generated before its start) and listens to the emission of new action events, showing their details at each update. 

When started, the monitor displays an output similar to the following:
\begin{lstlisting}[
    style=console,
    basicstyle=\fontsize{7}{8}\selectfont\ttfamily
]
 2018-06-12 11:57:57  99 deliverers[13] 1  X: 1117, Y: 257
 2018-06-12 12:10:31 102 deliverers[35] 1  X: 1283, Y: 677
 2018-06-12 12:09:53 101 deliverers[19] 1  X: 448,  Y: 407
 2018-06-12 12:15:39 103 deliverers[4]  1  X: 498,  Y: 261
 2018-06-12 11:48:52 100 deliverers[16] 1  X: 1015, Y: 363
 2018-06-12 13:07:01 100 deliverers[16] 2  X: 950,  Y: 300
\end{lstlisting}

Each line is an action performed by an agent on a shipment, The first four space-separated words are the timestamp of the actions inside the simulation (not in the current day). Then come the identifier of the shipment, the identifier of the agent specified in the simulation, the type of the action (1 for pick ups, 2 for deliveries), and finally the X-Y location where the action occurred. This allows users to clearly see what is stored in the blockchain.

\section*{Results and discussion}\label{sec:conclusion}

This section sums up what has been achieved in this report, shows our preliminary results and points out the downsides and the improvements that could be made.

\subsection*{Enforcing a shared shipment tracking system}

The simulation described earlier allows various agents to share the same blockchain network in order to store information on shipments they operate. This has two main benefits.

\paragraph{Shipment's single virtual identity and data availability}
Most of the time, companies involved in supply chain processes do not share the same information system. This results in information replication and/or information silos, \ie, tracking data are separated across the parties and one cannot access others' information directly.
With blockchain technology, it is possible to provide a shared and trustworthy information system (which is not managed by a single entity) so that data is available to everyone at each step, therefore improving transparency for businesses as well as consumers.

\paragraph{Agents' nature agnosticism}
Using this system, it is very easy to add new agents from new companies to participate in the shipping. 
All it takes is the creation of an account for each new agent, since agents' transactions are dealt with in the same way, whatever their company affiliation or their nature (deliverer or transporter in this case). In fact, agents need not even to be human beings, they could be IoT devices operating shipments automatically inside a hub. All they require is an Ethereum account.
Clearly, if a company or a shipment requires specific verification steps, the system could be adapted so that additional smart contracts might be deployed to enforce the required controls.

\subsection*{Blockchain size and data replication}

\begin{figure}
    \begin{center}
        \subfloat[Blockchain size as a function of time.]{
            \includegraphics[width=0.9\linewidth]{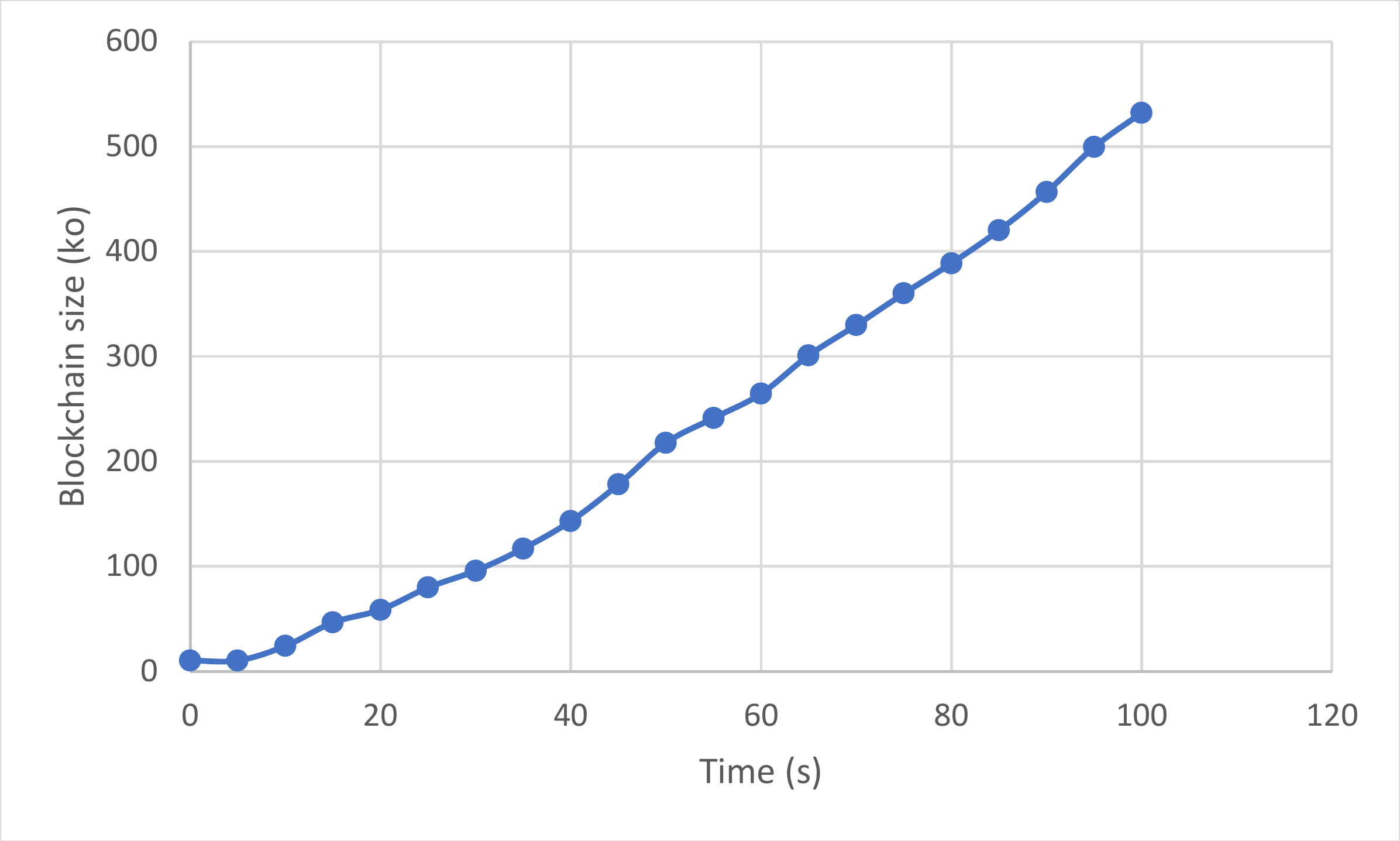}
            \label{subfig:bc-size-time}
        }
        
        \subfloat[Blockchain size as a function of the number of actions]{
            \includegraphics[width=0.9\linewidth]{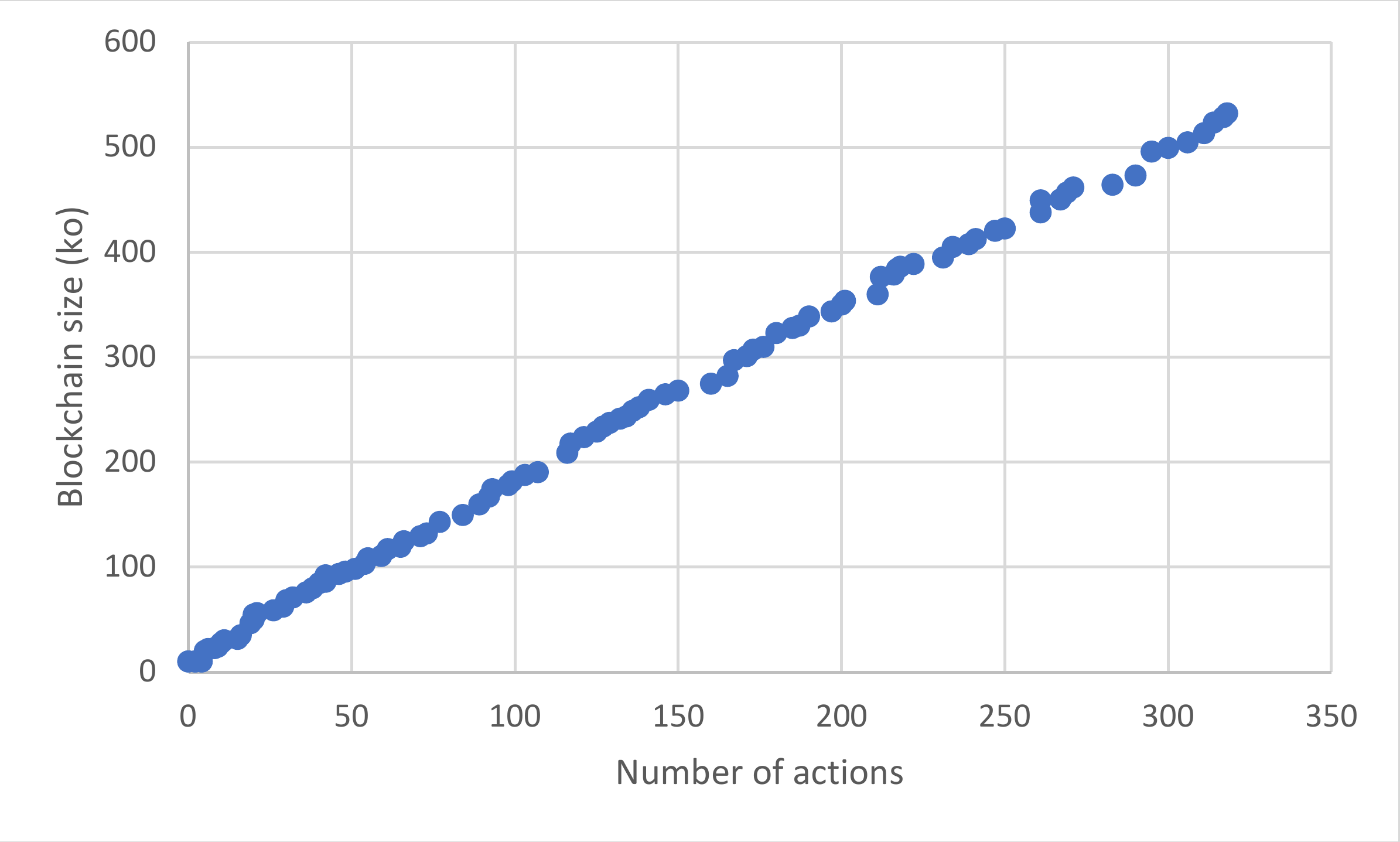}
            \label{subfig:bc-size-nb-actions}
        }
    \end{center}
    \caption{Blockchain size analysis.}
    \label{fig:bc-size-analysis}
\end{figure}

Illustratively, we ran the simulation for 100 seconds, which corresponds to about 4 hours in the simulated city, and we monitored the evolution of the blockchain size.
Figure \ref{fig:bc-size-analysis} shows the evolution of the blockchain size as a function of time (\ref{subfig:bc-size-time}), and as a function of the number of actions (\ref{subfig:bc-size-nb-actions}). After 100 seconds, the blockchain size reaches 532 KB, which corresponds to 318 actions sent by the AnyLogic agents.
Since the blockchain size evolves linearly against the action number, we can deduce that a transaction containing an action takes 1.6 KB in average.  

For large shipping companies like Amazon or UPS, this could mean that several thousands of gigabytes of data are potentially added every single day. 
Although it might not be an issue using current storage technologies, it is important to mention that every full node in a blockchain network stores the same blockchain, and that the security of blockchain networks relies on the fact that every full node must have a copy of it. Since their number
is expected to be quite high (e.g., in the Bitcoin network, there are currently about 10,000 full nodes \cite{bitnodes}), this means that there might be a huge amount of redundant information stored.

This problem might be tackled by storing the actual data in a regular distributed database instead of directly in the blockchain. To ensure the integrity and security of the data, the hashes of each action might be stored in the blockchain. This means that transactions would have a fixed size (due to the hash). Additionally, if the data volume ever became too large, one could decide to reinitialize the blockchain without losing the actual data. 

However, this implies that an entire external system must be set up and managed (in addition to the blockchain network) and most likely by a single entity, which might seem contradictory since the blockchain principle is that there is no need for central authorities or intermediaries. To avoid this last issue, note that it might be possible to implement this external database system in a P2P way, where each blockchain node would also be a database node, and where data is spread all over the nodes, with a minimal but necessary replication for security, availability and resilience purposes, like in BigchainDB \cite{Mcconaghy2016}.

\subsection*{Access control and confidentiality}

In our simulation, any agent connected to the network is able to send transactions and therefore provide tracking information about shipments. Each agent also has access to all the actions that have been or will be performed on the shipment. Although it is actually one of the main benefits of blockchain technology, this might also be quite dangerous. Any malicious agent could potentially send fake information to the network for various reasons (\eg, theft, corrupting sensible data, overloading the network), and could also access information he should not. 

This issue can be overcome by deploying smart contracts applying various access controls \cite{Zhang2018,hofman2017,hofman2017prez}, depending on the agent's address (which is directly linked to its public key, so it is not possible to fake it, as it would also fake the transaction signature). These smart contracts could, for example, allow only an updatable list of account addresses to append actions. However, although this solution protects against malicious write attempts, it does not resolve the issue of potential forbidden read access. Here, only two options appear possible currently: 
\begin{inparaenum}[1)]
\item encrypt sensible data and distribute the keys to the agents,
\item store data in an external system and only save data hashes in the blockchain (as we mentioned in the previous section).
\end{inparaenum}

\subsection*{Performance}

One of the main problem of the Ethereum network is that it supports a very low transaction rate (a few dozen per second on the main net, against 45,000 for Visa).
This can be a real issue if all companies use the same network, as it could be recommended by the Physical Internet paradigm, since hundreds of thousands transactions might occur at the same time across the world. However, other blockchain networks might be used to resolve this problem. For example, \emph{EOS.IO} is a blockchain structure which \textquote{has the potential to scale to millions of transactions per second} (from \url{https://eos.io/faq}).


\hfill

\lettrine{I}n this paper, we have illustrated that blockchain technology, with the IoT support for automated data collection, fits completely with the essence of the Physical Internet, and described its very first application to hyperconnected logistics. Despite some minor issues which will be the subject of future works, the technology is ready. Since such an infrastructure is expected to be used by many different entities, like companies, governments and consumers, one of the upcoming challenges is to establish standards and to design an architecture that would support and suit all the parties. 



\begin{thebibliography}{10}

\bibitem{Atzori2010}
L.~Atzori, A.~Iera, and G.~Morabito.
\newblock {The Internet of Things: A survey}.
\newblock {\em Computer Networks}, 54(15):2787--2805, oct 2010.

\bibitem{crainic2016}
T.~G. Crainic and B.~Montreuil.
\newblock Physical {Internet} {Enabled} {Hyperconnected} {City} {Logistics}.
\newblock {\em Transportation Research Procedia}, 12:383--398, 2016.

\bibitem{Halle2018}
S.~Hall{\'{e}}, R.~Khoury, Q.~Betti, A.~El-Hokayem, and Y.~Falcone.
\newblock {Decentralized enforcement of document lifecycle constraints}.
\newblock {\em Information Systems}, 74:117--135, may 2018.

\bibitem{hofman2019}
W.~Hofman and C.~Brewster.
\newblock The {Applicability} of {Blockchain} {Technology} in the {Mobility}
  and {Logistics} {Domain}.
\newblock In {\em Towards {User}-{Centric} {Transport} in {Europe}:
  {Challenges}, {Solutions} and {Collaborations}}, Lecture {Notes} in
  {Mobility}, pages 185--201. Springer International Publishing, 2019.

\bibitem{hofman2017}
W.~Hofman, J.~Spek, and C.~Brewster.
\newblock Applying blockchain technology for hyperconnected logistics.
\newblock In {\em 4th International Physical Internet Conference}, Graz,
  Austria, 2017.

\bibitem{hofman2017prez}
W.~Hofman, J.~Spek, and C.~Brewster.
\newblock Applying {Blockchain} {Technology} for {Hyperconnected} {Logistics},
  2017.
\newblock Presented at the 4th International Physical Internet Conference,
  Graz, Austria. Available at
  \url{https://www.pi.events/IPIC2017/sites/default/files/IPIC2017-Workshop-2.2_presentations.pdf}.
  Accessed: 2019-04-02.

\bibitem{Kaboudvand2018}
S.~Kaboudvand, B.~Montreuil, S.~Buckley, and L.~Faugere.
\newblock {Hyperconnected Megacity Logistics Service Network Assessment: A
  Simulation Sandbox Approach}.
\newblock 2018 IISE Annual Conference, May 2018.

\bibitem{Kelepouris2007}
T.~Kelepouris, K.~Pramatari, and G.~Doukidis.
\newblock {RFID}-enabled traceability in the food supply chain.
\newblock {\em Industrial Management {\&} Data Systems}, 2007.

\bibitem{Korpela}
K.~Korpela, J.~Hallikas, and T.~Dahlberg.
\newblock {Digital Supply Chain Transformation toward Blockchain Integration}.
\newblock In {\em Hawaii International Conference on System Sciences (HICSS)},
  2017.

\bibitem{bootnodes}
F.~Lange.
\newblock Setting up private network or local cluster, 2017.
\newblock Available at
  \url{https://github.com/ethereum/go-ethereum/wiki/Setting-up-private-network-or-local-cluster}.
  Accessed: 2018-11-20.

\bibitem{Li2011}
L.~Li.
\newblock {Application of the Internet of Things in green agricultural products
  supply chain management}.
\newblock In {\em Proceedings - 4th International Conference on Intelligent
  Computation Technology and Automation, ICICTA 2011}, 2011.

\bibitem{Lu2017}
Q.~Lu and X.~Xu.
\newblock {Adaptable Blockchain-Based Systems: A Case Study for Product
  Traceability}.
\newblock {\em IEEE Software}, 34(6):21--27, nov 2017.

\bibitem{Madhwal2017}
Y.~Madhwal and P.~B. Panfilov.
\newblock {Industrial Case : Blockchain on Aircraft ' s Parts Supply Chain
  Management.}
\newblock In {\em American Conference On Information Systems 2017 Workshop On
  Smart Manufacturing Proceedings}, 2017.

\bibitem{Mcconaghy2016}
T.~Mcconaghy, R.~Marques, A.~M{\"{u}}ller, D.~{De Jonghe}, T.~Mcconaghy,
  G.~Mcmullen, R.~Henderson, S.~Bellemare, and A.~Granzotto.
\newblock {BigchainDB: A Scalable Blockchain Database (DRAFT)}.
\newblock {\em BigchainDB}, 2016.

\bibitem{PI}
B.~Montreuil.
\newblock {Toward a Physical Internet: meeting the global logistics
  sustainability grand challenge}.
\newblock {\em Logistics Research}, 3(2-3):71--87, may 2011.

\bibitem{Montreuil2018}
B.~Montreuil, S.~Buckley, L.~Faugere, R.~Khir, S.~Derhami, M.~Benoit,
  B.~Shannon, F.~Louis, K.~Reem, and D.~Shahab.
\newblock {Urban Parcel Logistics Hub and Network Design: The Impact of
  Modularity and Hyperconnectivity}.
\newblock In {\em 15th IMHRC Proceedings}, chapter~19. Savannah, Georgia, USA,
  2018.

\bibitem{morley2016}
M.~Morley.
\newblock How {IoT} {Enables} a {Hyper} {Connected} {Supply} {Chain}, Oct.
  2016.
\newblock Available at
  \url{https://blogs.opentext.com/how-iot-enables-a-hyper-connected-supply-chain/}.
  Accessed: 2019-04-02.

\bibitem{Sallez2016}
Y.~Sallez, S.~Pan, B.~Montreuil, T.~Berger, and E.~Ballot.
\newblock {On the activeness of intelligent Physical Internet containers}.
\newblock {\em Computers in Industry}, 81:96--104, sep 2016.

\bibitem{Tian2017}
F.~Tian.
\newblock {A supply chain traceability system for food safety based on HACCP,
  blockchain {\&} Internet of things}.
\newblock In {\em 14th International Conference on Services Systems and
  Services Management, ICSSSM 2017 - Proceedings}, 2017.

\bibitem{Toyoda2017}
K.~Toyoda, P.~T. Mathiopoulos, I.~Sasase, and T.~Ohtsuki.
\newblock {A Novel Blockchain-Based Product Ownership Management System (POMS)
  for Anti-Counterfeits in the Post Supply Chain}.
\newblock {\em IEEE Access}, 5:17465--17477, 2017.

\bibitem{bitnodes}
A.~Yeow.
\newblock Bitnodes.
\newblock Available at \url{https://bitnodes.earn.com/dashboard/}. Accessed:
  2018-11-20.

\bibitem{Zhang2018}
Y.~Zhang, S.~Kasahara, Y.~Shen, X.~Jiang, and J.~Wan.
\newblock {Smart Contract-Based Access Control for the Internet of Things}.
\newblock {\em IEEE Internet of Things Journal}, 2018.

\end{thebibliography}


\vskip 20pt

\noindent
{\it
\textbf{Quentin Betti} is a doctoral student at Université du Québec à Chicoutimi. \textbf{Raphaël Khoury} and \textbf{Sylvain Hallé} are professors at Université du Québec à Chicoutimi, Canada; Pr.\ Hallé is also the Canada Research Chair on Software Specification, Testing and Verification. \textbf{Benoit Montreuil} is Professor in the Stewart School of Industrial and Systems Engineering at the Georgia Institute of Technology, where he is the Coca-Cola Chair in Material Handling and Distribution, Director of the Supply Chain \& Logistics Institute, and Director of the Physical Internet Center.
}


\end{document}